\documentstyle[preprint,prl,aps]{revtex}

\begin{document}
\draft

\title{Are there  basic laws of quantum information processing?}

\author{Micha\l{} Horodecki \cite{poczta1}}

\address{Department of Mathematics and Physics\\
 University of Gda\'nsk, 80--952 Gda\'nsk, Poland}

\author{Ryszard Horodecki\cite{poczta2}}

\address{Institute of Theoretical Physics and Astrophysics\\
University of Gda\'nsk, 80--952 Gda\'nsk, Poland}

\maketitle

\begin{abstract}
We prove within the standard quantum formalism without reduction postulate
that the no-cloning theorem and the principle
of no-increasing of entanglement under local actions and one-way classical
communication are equivalent. We argue that the result is a manifestation 
of more general principles governing quantum information processing analogous
to the thermodynamical laws. 
\end{abstract}

\newpage
\section{Introduction}
The recent development of quantum information theory shows that the no-cloning
theorem \cite{clone} can be considered as a basic principle of quantum
information processing (QIP). It  provides fundamental bounds for capacities
of quantum channels. Quite recently
Bennett {\it et al.} \cite{erasure} calculating the capacities of
simple quantum channels supplemented
(or not) by classical ones, used the theorem as a basic tool for providing
upper bounds for the capacities. For example, this allowed to show that the
so-called quantum erasure channel with probability $\epsilon=1/2$ of erasing
has the quantum capacity equal to zero i.e. cannot be used for reliable
transmission of quantum information.

On the other hand the no-cloning theorem has been also used  in the context of
entanglement processing concerning sending quantum information by means
of teleportation \cite{telep}. Recently Bennett {\it et al.} \cite{pur}
introduced a concept of obtaining pure singlets needed for
faithful teleportation from
mixed states by means of local quantum operations
and classical communication (LCC operations).
This is called distillation (or purification) protocol. It
appears \cite{huge} that the no-cloning theorem allows in some cases
to obtain the
bounds for the maximal asymptotic yield of produced singlets (called
distillable entanglement) for a
given mixed state. This becomes more clear, in light of a general connection
between quantum channel capacities and distillable entanglement of mixed
bipartite states \cite{huge}. It involves
the fact that many properties of a quantum channel
(being in  general a completely positive trace preserving map) can be
conveniently described by properties of mixed state produced by sending part
of maximally entangled bipartite state through it \cite{Nielsen}. Thus
considering entanglement can be helpful in the investigation of channels
and vice-versa. This can be illustrated by the fact that it is only
the possibility of
sending quantum information by means of entangled state and classical
communication (i.e. teleportation) which can make the capacity of
a quantum channel supplemented by two way classical channel to be
greater than in the case of the quantum channel alone.

Now let us note that the entanglement processing involves another basic
principle.  It says that  entanglement cannot
increase under local action and classical communication (we will refer to it
as to ``no-increasing of entanglement'')
\footnote{
By no-increasing of entanglement under LCC operations we mean that
if the output ensemble is $\{p_i, \varrho_i\}$ at
the input state $\varrho$ then  
(i) $\sum_ip_iE(\varrho_i)\leq E(\varrho)$ (i.e. that the average input
entanglement cannot exceed the entanglement of input state) \cite{huge} 
where E is a given entanglement measure.}
This principle, being proved as a theorem for a particular entanglement
measure (entanglement of formation) \cite{huge}, can be treated as a
postulate for any good entanglement measure \cite{huge,termo,Knight}. There
was also made some attempt to consider it as an analogue of second law of
thermodynamics \cite{termo}.

Now a fundamental question arises. Is there a connection between the two
principles: the first
concerning the real and the second  - ``virtual'' quantum information
\footnote{The term ``virtual'' information was introduced in Ref.  
\cite{Cerf}.} due to  entanglement ?
Note that the possible direct connection could be inferred from the recent
result of Bu\v{z}ek {\it et al.} \cite{Buzek2},
who used the imperfect quantum copiers \cite{Buzek1} locally to produce two
entangled   states from one input entangled one. Thus imperfect cloning allows
for imperfect ``broadcasting inseparability''. The fact that the entanglement
of the joint output state cannot be greater than the initial one can be
viewed as a reflection of imperfection of the copying machines. This suggests
that this is just impossibility of cloning which is responsible for the fact
that one cannot increase entanglement by LCC operations. Consequently, one can
ask the converse question: is it that  the no-increasing of entanglement
principle implies the no-cloning theorem? More generally, one can ask: are
there generic principles of quantum information processing?

The aim of this Letter is to give at least partial answers to the above
questions. In sec. 2 we prove the equivalence between no-cloning theorem
and the principle of no-creating of entanglement by means of
local quantum operations and  one-way classical communication. In sec. 3
we consider the equivalence theorem in the context of the recent
results in quantum information theory. It leads us to the two principles of
quantum information processing being analogous to the first and second
principles of thermodynamics. In the last section we discuss physical
consequences  of the principles.

\section{The equivalence theorem}
In this section we will show that the no-cloning theorem and the principle of
no-increasing of entanglement (in a bit weaker formulation than usual ones)
are equivalent. We start with the following formulations

\begin{enumerate}
\item  No-cloning theorem: {\it There does not exist a quantum operation
 which clones every input pure state \footnote{ {\rm
The no-cloning theorem in the strongest version  rules out the possibility
of cloning even two nonorthogonal states, see Ref. \cite{strong}}}.}

\item  No-creating \footnote{
{\rm According to our formulation of the principle the term ``no-creating''
seems to be here more appropriate than ``no-increasing''.
}}
of entanglement principle: {\it
There does not exist an LCC operation which produces singlet pairs from
product input state.}
\end{enumerate}

In the above formulations we allow the cloning or singlet-creating machines
to produce output with any desired but not necessarily perfect accuracy.
Note that the principle (2) says simply that the distillable entanglement of
product state vanishes which is of course a mathematical fact. Note also that
in the above formulation we do not employ any particular entanglement measure.
Of course we can divide the LCC operation into the ones involving one- or
two-way classical communication.
Below we restrict ourselves to the no-one-way-entanglement-creation principle
in the form

$2'$.\  {\it There does not exist an LCC
operation with one-way classical channel
which produces singlet pairs from product input state.}

To be precise, by a quantum operation (in short: operation) on a system A
we mean a completely positive trace preserving map
$\Lambda_A:{\cal B}({\cal H}_A)\rightarrow{\cal B}({\cal H}_A)$ where
${\cal H}_A$ is the Hilbert space associated with the system A and
${\cal B}({\cal H}_A)$ denotes the set of linear operators acting on
${\cal H}_A$. Any such a map is of the form
\begin{equation}
\Lambda_A(\varrho)=\sum_iV_i\varrho V_i^\dagger\quad {\rm with}
\sum_iV_i^\dagger V_i=I_A,
\end{equation}
where $I_A$ denotes identity operator on ${\cal H}_A$.
It is known \cite{CP} that the map is also of the form  
\begin{equation}
\Lambda_A(\varrho)={\rm Tr}_C[U\varrho\otimes\omega U^\dagger], 
\end{equation}
where $\omega$ is a state on the additional system C and U is unitary
transformation on the joint system A+C.

By an LCC operation with one-way classical channel  on the compound system A+B
we mean an operation  which can be written in the form
\begin{equation}
\Lambda_{AB}(\varrho) = \sum_i \hat{I}_A\otimes
\Lambda_B^i(V_A^i\otimes I_B\varrho
{V_A^i}^\dagger \otimes I_B),
\label{lccone}
\end{equation}
where the $\{V^A_i\}$ is a partition of unity  on the system A,
$\Lambda_B^i$ are operations on the system B and $\hat{I}_A$ is identity
operation on the system A (i.e. it is a superoperator).

Now we  will prove the following

{\bf Theorem.-} {\it The no-cloning theorem  {\rm (1)} and the principle of
no-one-way-creating of entanglement  {\rm ($2'$)} are equivalent.}

{\bf Proof.-}
We will prove that the negations of the principles 1 and $2'$ are equivalent.
We will usually deal  with  exact cloning or singlet-creating
machines. Then the more general case  (where any desired but not exact
accuracy is required) will be also true due to compactness of bounded sets
in finite-dimensional linear spaces.

Suppose now that given an initial product state $P_A\otimes P_B$
Alice and Bob are able to produce entangled state $P^{ent}_{AB}$
by local quantum operation and one-way classical communication i. e.
by an operation of the form (\ref{lccone}) 
where  $\{V^A_i\}$ is the partition of unity corresponding to Alice's
generalized measurement and $\Lambda_B^i$ denotes Bob's operation  performed
if the i-th outcome is obtained by Alice.
Consider then an unknown state $P_C$ on an additional system $C$ in Alice's
laboratory. After performing the operation $\Lambda_{AB}$ the state $P_C$ can
be teleported to Bob via the state $P^{ent}_{AB}$. The operation of
teleportation is of the same form as $\Lambda_{AB}$
\begin{equation}
\Lambda^{tel}_{CAB}(\varrho)= \sum_i\hat{I}_{CA}\otimes\tilde{\Lambda}_B^i(
\tilde
{V}_{CA}^i\otimes I_B\varrho \tilde{V}_{CA}^{i\dagger} \otimes I_B).
\end{equation}
Combining the two actions on the system C+A+B i.e. $I_C\otimes \Lambda_{AB}$
and $\Lambda^{tel}_{CAB}$ we obtain the joint action
\begin{equation}
\Lambda_{CAB} (\varrho)= \sum_{ij} \hat{I}_{CA}\otimes\Lambda_B^{ij}(
W_{CA}^{ij}\otimes I_B\varrho  {W}_{CA}^{ij\dagger} \otimes I_B),
\end{equation}
where $\Lambda_B^{ij}=\tilde{\Lambda}_B^j\Lambda_B^i$,
$W^{ij}_{CA}=\tilde V_{CA}^j (I_C\otimes V^i_A)$. Then we have
for any unknown state $P_C$
\begin{equation}
{\rm Tr}_{CA} \left[\sum_{ij}(W^{ij}_{CA} P_C\otimes P_A
 W^{ij\dagger}_{CA})
\otimes \Lambda_B^{ij}(P_B)\right]=P_C.
\end{equation}
Now it follows that  the operation
\begin{equation}
\Lambda_{CAB_1B_2\ldots B_n} (\varrho)= \sum_{ij} \hat{I}_{CA}\otimes
\left(\otimes_{i=1}^n\Lambda_{B}^{ij}\right)(
W_{CA}^{ij}\otimes I_B\varrho  W_{CA}^{ij\dagger} \otimes I_B)
\end{equation}
will clone the unknown state $P_C$.
To see it note that  for any operators $A_i$ and $B_i$  with
${\rm Tr} B_i=1 $ we have
\begin{equation}
{\rm Tr}_{12} \left[\sum_i A_i \otimes B_i \otimes B_i\right]=
{\rm Tr}_{13} \left[\sum_i A_i \otimes B_i \otimes B_i\right]=
{\rm Tr}_{1} \left[\sum_i A_i \otimes B_i \right].
\end{equation}
Then for any state $P_C$
\begin{equation}
{\rm Tr}_{CA} \left[\Lambda_{CAB_1B_2\ldots B_n} \left(P_C\otimes P_A\otimes
(\otimes_{i=1}^n
P_{B} )\right)\right]= \otimes_{i=1}^n P_C.
\end{equation}
Thus from no-cloning theorem follows the no-entanglement-creating one.

Suppose now conversely, that we are able to clone quantum states i.e. that
there exists an operation $\Lambda_{AC_1\ldots C_n}$ and state
$\otimes_{i=1}^n P_C$ such that for any unknown state $P_A$
\begin{equation}
\Lambda^{copy}_{AC_1\ldots C_n}(P_A\otimes(\otimes_{i=1}^{n} P_C))=
\otimes_{i=1}^{n+1} P_A
\end{equation}
Performing this operation  Alice obtains $n+1$ (arbitrarily good)
copies of the initial state $P_A$. Then there exists a measurement $\{P_i\}$
on the system $AC_1\ldots C_n$ such that
to any outcome $i$ Alice can ascribe a guessed state $\tilde P_i$ with
$\sum_i p_i \tilde P_i$ closely approximating $P_A$ for large n; $p_i$ denote
the probabilities of the corresponding outcomes (for optimal measurement in
the spin--$1\over2$ case see \cite{PopMas}).
The two actions of Alice -- copying and measurement -- can be treated jointly
as a generalized measurement performed on the system A
given by partition of unity $\{V^i_A\}$. Now obtained an outcome $i$ Alice
sends it
classically to Bob, who subjects his initial state $P_B$ to an operation
$\Lambda_B^i$ to obtain $\tilde P_i$
\begin{equation}
\Lambda_B^i(P_B)=\tilde P_i
\end{equation}
Combining actions of Alice and Bob we obtain joint operation $\Lambda^n_{AB}$
which is of course LCC operation with one-way classical channel
\begin{equation}
\Lambda^N_{AB} (\varrho)=\sum_i I_A \otimes \Lambda_B^{ni}(V_A^{ni}\otimes I_B
\varrho V_A^{ni\dagger}\otimes I_B)
\end{equation}
and which has the property that for all $P_A$ and for n large enough
${\rm Tr}_A \left[\Lambda^n_{AB}(P_A\otimes P_B\right]$
closely approximates $P_A$.
Now it suffices to show that entanglement can be created by means of a map
$\Lambda_{AB}$ for which there holds exact equality
\begin{equation}
{\rm Tr}_A \left[\Lambda_{AB}(P_A\otimes P_B\right]=P_A
\label{exact}
\end{equation}
To create entanglement
Alice prepares entangled state $P^{ent}_{AC}$ on the system $A+C$, where $C$
is ancilla, and then Alice and Bob apply the operation $\Lambda_{AB}$.
Now we have to show that if the latter map flips any state $P_A$ onto Bob's
system, then also the entanglement will be suitably swapped. Equivalently one
can show that  $\Lambda_{AB}$ followed by the flip operator
($\Lambda_V(P\otimes Q)=Q\otimes P$) will leave entanglement A and C
undisturbed. Consider then the map $\Lambda_A(\varrho)={\rm Tr}
(\Lambda_V \Lambda_{AB}(\varrho\otimes P_B))$. By the property
(\ref{exact}) such a map must be identity, so acting on half of entangled
state on the system A+C will leave it undisturbed
\begin{equation}
(\Lambda_A\otimes I_C)(P_{AC}^{ent})=P_{AC}^{ent}.
\end{equation}

\section{Possible implications of the equivalence theorem:
principles of quantum information processing}

Now it is natural  to ask what are  physical implications of the equivalence
theorem. To give answer to this question, we adopt the view according to which
quantum information is a physical quantity \cite{Ann,Landauer}
similarly as energy in  thermodynamics. 
Below we will consider the physical meaning of the equivalence theorem in this
spirit.

{\it A. First principle.}
In the proof of the theorem we have utilized the fact that if a physical
operation sends any state undisturbed then it will also swap entanglement
without destroying it. To understand it, note that if we
work within the orthodox quantum formalism  (without
the von Neumann projection postulate) then acting on one part of
entangled system, we have no way to {\it annihilate} the entanglement. The
latter can change only by means of interacting of the {\it both} entangled
subsystems. The only way we can disturb it is to entangle the considered
subsystem with another one. Then the entanglement will not of course vanish
but it will spread over all the three subsystems. Now if the action leaves
any state of the subsystem undisturbed, then it obviously does not entangle it
with any other one. Consequently, the entanglement of the initial system
remains unchanged. The same concerns the operation which sends any state
without any disturbance to another system. Then it will also perfectly swap
entanglement.

Note that the present reasoning can be viewed as utilizing of some
implication of {\it conservation of quantum information} (which can be called
first principle of QIP in analogy to that of thermodynamics).
It can be formulated as follows

\noindent
{$I'$. \it The entanglement of the compound system does not change under
unitary processes on one of the subsystems.}

Indeed the equivalence theorem strongly confirms the recently more
and more common view that entanglement is simply a form of quantum
information. Then the rules of processing of quantum information can
always be formulated in two ways. The above formulation expresses the principle
of conservation of information in the language of virtual quantum information
represented by half of entangled state.
We can express it also in terms of real quantum information,  i.e. the one
associated with {\it unknown parameters} of state of a particle.
Consider for this purpose a  system in an unknown state
representing a qubit of quantum information and  another system in an
established state (hence representing no quantum information). Now, if the
unitary transformation over the whole system is performed, one can observe
how the information is spread over the system. As there was one qubit of
initial amount of information and the information of the whole system cannot
change, after evolution we are left also with one qubit of information.
However, the latter is in general no longer contained in the first subsystem,
but part of it could be flipped onto the second one or  changed into
quantum entanglement of the whole system. Then the alternative formulation of
the first principle of QIP can be as follows

\noindent
{I. \it For a compound quantum system  the sum of information
contained in the subsystems and the information contained in  entanglement
is conserved in unitary processes}.

As an  illustration of the above
principle can  serve the observation \cite{Buzeknet,GisinMas}
that during quantum copying process the information is not lost
but can be found in entanglement of the output qubits and copying machine.
Note that here the entanglement represents real quantum information, as
it contains the {\it unknown} parameters of the superposition of the
input qubit,
while the virtual information is identified with entanglement {\it itself} or,
more precisely, half of a system being in a {\it known}, established entangled
state.

{\it B. Second principle.}
Having formulated the information conservation principle in the two ways we are
in position to interpret the result of the previous section. The two
equivalent statements $1$ and $2'$ are formulations of  {\it the same}
restriction for quantum information processing in terms of quantum
{\it real} and {\it  virtual} information respectively. Following the proof,
one can see that the no-cloning theorem says in fact that we cannot send
quantum real
information through classical channel. As we work within quantum formalism
(where there are no ``classical'' channels)
the above means that it is not possible to flip a qubit from one system to
another by some sort of quantum operations on the whole system i.e. by
the operations of the form (\ref{lccone}),
where we think about them  as if they were  a {\it part of unitary
transformation} over the system of interest plus some unobserved ancilla.
Then the equivalence
theorem says that this impossibility is equivalent to impossibility of
swapping half of entangled system onto another one by means of such operations.
The latter impossibility we have called in short no-creating entanglement
principle. However it must be emphasized here, that it is {\it not} a
consequence of information conservation. The latter implies merely
that acting on {\it one} of the
system we cannot create or annihilate entanglement. Now the map (\ref{lccone})
acts on both the subsystems. Thus the equivalent no-cloning or no-creation
of entanglement principles constitute a different physical law which says that
neither real nor virtual quantum information can be transmitted by such type
of channels. One can consider stronger prohibition: quantum information cannot
be transmitted by a so-called separable superoperator \cite{Knight,Rains}
\begin{equation}
\Lambda(\varrho)=\sum_i V_A^i\otimes V_B^i \varrho V_A^{i\dagger}\otimes
V_B^{i\dagger}.
\end{equation}
Under the consideration one can formulate the  second principle of QIP
as follows

\noindent
{II. \it One cannot spread quantum real information from an initial system onto
larger number of systems without changing part of it into entanglement.}

To
formulate it in terms of virtual information, imagine that two systems A and B
are entangled, and we can act only on the system A and some additional system
C. Then it is impossible to make C entangled with B in such a way that all
entanglement of the system ABC is contained in the subsystems AB and AC.
This can be formulated almost identically as in the case of  real 
information, if we recall the idea of ``entangled entanglement'' \cite{Zei}.
Indeed, similarly as the quantum real information is partly converted
into entanglement,  the lacking part of virtual information which is not
contained in the pairs AB and AC, but only in  the whole system ABC, may be
thought as being converted into ``entanglement of entanglement''. In this
way we have come to the suitable formulation the second principle of QIP
in terms of virtual information

\noindent
{$II'$. \it It is impossible to spread
virtual information  without changing part of it into entangled entanglement.}

It can be seen that the second  principle of QIP implies the no-cloning
principle. As the latter has been  proved to be equivalent to no-creation of
entanglement, we see that our formulations of the second principle are closely
related to the one proposed by Popescu and Rohrlich \cite{termo} who postulated
no-increasing of entanglement as an analogue to the thermodynamical law.

\section{Discussion and conclusion}

That the second principle of QIP may play the  role analogous to the  second 
law of thermodynamics we can see from the fact that in our formulation 
the former implies the
no-cloning theorem. As it was mentioned, the latter  provides fundamental
limits for capacities of quantum channels. Analogously,
the impossibility of sending information by a channel constituted by
local quantum operations and two-way classical communication
(also implied by the second principle) provides bounds
for capacities of quantum channels supplemented with two-way classical
channels. This is analogous to the fact that it is just the second law of
thermodynamics what determines the  limits for efficiency of cyclic 
machines acting
between two heat reservoirs. Now the parameters of noisy channels can be
considered as a counterpart of the temperatures   $T_1$ and $T_2$ 
of reservoirs, while the error-correcting protocols would correspond to the
thermodynamical processes.  The central problem in the quantum
information theory of  finding the optimal protocol for a given noisy
channel can  be then compared  with the problem of finding the engine 
(i.e. the cyclic process) with the best efficiency for given 
temperatures $T_1$ and $T_2$, what was done by Carnot.

Let us now consider briefly the problem of the ``time arrow'' in the 
context of the
second law of QIP. It may be objected that the latter cannot
be suitable counterpart of the thermodynamical principle as
there is no irreversibility in the used formalism.
However, the discussed
notions as e.g. quantum information, can be considered independently of
quantum formalism. For example, if the quantum computer were realized in 
laboratories,
the quantum information would not be an abstract notion any more, but it would 
become a very concrete physical quantity. Then we can observe an  analogue
of the ``arrow'' which is permanently ascribed to the second law of
thermodynamics.
Namely in the real world we still observe the
decoherence \cite{Zurek}. The latter is nothing else but in practice
irreversible  entangling of a number of systems. Thus, although the
decoherence is, in standard quantum formalism,  logically reversible,
in real world there is  a tendency that quantum information prefers to be
spread over many systems, and necessarily changes into {\it uncontrolled}
entanglement.
But what is uncontrolled entanglement? According to our analogy,
it is simply ``information heat flaw''.
Indeed, the thermodynamical heat flaw means  in fact increase of internal
energy of uncontrolled reservoir.

Here a question arises: where is the place of the quantum entropy in the
above framework? This is a fundamental problem, as the
second law of thermodynamics is inherently connected with the notion
of entropy.
But in our approach the entropy of the  universe is constant, as we
consider the latter to be a closed system, to keep the conservation
information principle valid
\footnote{This is consistent with the concept of unitary information field
\cite{Ann}, which rests on the assumption that the notion of information is a
fundamental category in the description of reality.}.
A possible way out is that
according to the interpretation of our formulation of second law of
QIP, one can divide the universe into the  elementary subsystems
and note that the tendency of increasing of entanglement between them makes
their entropies to become larger and larger. Then, in result, the sum
of {\it partial} entropies increases. It can be viewed as a suitable 
analogue of the statement in the standard thermodynamics that ``the 
entropy of the world increases''.

It should be emphasized here that the quantum
information has no its {\it own} dynamics, but it is carried by the particles
the  motion of which is
governed  by a dynamics which  is unavoidably connected with the concept of
energy. In fact there are two
levels in nature:   energetic (thermodynamical) and informational
(logical). They are mutually coupled because the dynamics starts with some
initial conditions which can be considered  an initial package of information.
Then it follows that the informational ``arrow'' is a reflection of the
thermodynamical one. This suggests, that the developed here analogy between
quantum informational and thermodynamical processes is not just a formal
one, but has a deeper physical justification.

Finally we have to discuss our results in the context of  the postulate
reduction of the wave function. In the mathematical proof and the
discussion in the previous section we have considered Alice's and Bob's
actions to be   unitary transformations over large system. This cannot be so,
if we treat reduction of wave packet as a real physical process.
Then the whole Alice-Bob action cannot be thought as a part of unitary
transformation and then the proof of  equivalence is not valid any more. Such a
situation arises when the observers, Alice and Bob, are not considered to
be a part of the quantum world but rather belong to another ``classical''
world \cite{Jadczyk}.
Then indeed, the wave packet reduction actually holds, the quantum
information is lost (entanglement is annihilated) and we must abandon the
information conservation principle.
One could ask now, when
the joint action of Alice and Bob can be viewed as a part of unitary
transformation? The answer is: only if we include the observers into the
quantum world.
Then entanglement cannot be annihilated by  any action which
does not affect both entangled systems.
In other words, to keep the quantum information conservation valid, one
is forced to work within the interpretations of quantum mechanics which
dispense with the postulate of  collapse of wave function
\footnote{
See in this context Refs. \cite{Everett,Bohm}.}.
This means that  the validity of the first  principle and the 
equivalence theorem appears to be interpretation-dependent.

To summarize, we have proved that within the quantum formalism without
reduction postulate the no-cloning theorem is equivalent
to the fact that one cannot produce entanglement from product state by means
of local quantum operations and  one-way classical communication. Then
we have postulated some principles  of QIP analogous to those in
thermodynamics. We have expressed  both the principles by means of real
and virtual information where the real quantum information is represented by
an unknown quantum state while virtual one -- by half of entangled system.
The  statements  proved to be equivalent appear now to constitute  the same
restriction
for quantum information processing expressed in the two different ways.
They appear to be a consequence of the second principle of QIP. The 
first principle,
being information conservation law, does not hold if the postulate of
reduction of wave packet is considered as real physical process.

The authors are grateful to Charles Bennett, Pawe\l{} Horodecki,
Martin Plenio and Vlatko Vedral for stimulating discussions and 
helpful comments.
This work is supported in part by Polish Committee
for Scientific Research, Contract No. 2 P03B 024 12.

\end{document}